\documentclass[%
reprint,
superscriptaddress,
 amsmath,amssymb,
 aps,
]{revtex4-1}

\usepackage[T1]{fontenc}
\usepackage{stix}
\usepackage{listings}

\usepackage{graphicx}
\usepackage{dcolumn}
\usepackage{bm}
\usepackage{hyperref}

\usepackage{xcolor}

\newcommand{\summary}[1]{}

\begin{document}
	
	\title{Wannier Diagram and Brown-Zak Fermions of Graphene on Hexagonal Boron-Nitride.}

	\date{\today}


	\author
	{
		T. Fabian, M.~Kausel, L.~Linhart, J. Burgd\"orfer, and F. Libisch$^*$,\\
		\normalsize{Institute for Theoretical Physics, TU Wien, 1040 Vienna, Austria, EU}\\
		\normalsize{$^{*}$Corresponding author; E-mail: florian.libisch@tuwien.ac.at}
	}

	\begin{abstract}
		The moir\'e potential of graphene on hexagonal boron nitride (hBN) generates a supercell sufficiently large as to thread a full magnetic flux quantum $\Phi_0$ for experimentally accessible magnetic field strengths.
		Close to rational fractions of $\Phi_0$, $p/q \cdot\Phi_0$, magnetotranslation invariance is restored giving rise to Brown-Zak fermions featuring the same dispersion relation as in the absence of the field. 
		Employing a highly efficient numerical approach we have performed the first realistic simulation of the magnetoconductance for a 250 nm wide graphene ribbon on hexagonal boron nitride using a full ab-initio derived parametrization including strain.
		The resulting Hofstadter butterfly is analyzed in terms of a novel Wannier diagram for Landau spectra of Dirac particles that includes the lifting of the spin and valley degeneracy by the magnetic field and the moir\'e potential. 
		This complex diagram can account for many experimentally observed features on a single-particle level, such as spin and valley degeneracy lifting and a non-periodicidy in $\Phi_0$.
	\end{abstract}
	
	\maketitle
	
	Ultraclean graphene in crystallographic alignment with hexagonal boron nitride (hBN) forms large-scale moiré patterns\cite{PhysRevLett.101.126804,Yankowitz2012}.
	The area $S \approx 165$ nm$^2$ of one supercell is large enough to reach the regime of one magnetic flux quantum $\Phi_0=h/e$ per moiré unit cell already at $B=23.5$~T.
	A plethora of fascinating phenomena emerge such as Hofstadter's butterfly, the fractal energy spectrum of the lattice in the magnetic field \cite{PhysRevB.14.2239,Rammal1985}, Weiss oscillations \cite{PhysRevB.75.125429,2106.11328} and Brown-Zak oscillations  \cite{PhysRev.133.A1038,PhysRev.134.A1602,PhysRev.134.A1607}, revivals of zero-field conductivity at large magnetic fields periodic in $1/B$. 
	Indeed, Hofstadter's butterfly could recently be experimentally observed \cite{Ponomarenko2013,Dean2013,Hunt1427,Yu2014,Wang1231,KrishnaKumar181,KrishnaKumar5135,Barrier2020}. 
	For exactly rational fractions $p/q \, \Phi_0$ of the magnetic flux quantum corresponding to magnetic field values $B_{p/q} = p/q\, \Phi_0/S$, the resulting Bloch wavefunction is strictly periodic and magneto-translational invariance is restored.
	Bloch's theorem predicts new quasi-particles, so-called Brown-Zak fermions (BZfs) \cite{PhysRev.133.A1038,PhysRev.134.A1602,PhysRev.134.A1607}, which travel, undeflected by the magnetic field, ballistically through the moiré superlattice \cite{Bloch1929,PhysRev.133.A1038,PhysRev.134.A1602,PhysRev.134.A1607,azbel1964energy,https://doi.org/10.1002/pssb.2220880243,PhysRevB.14.2239,Streda_1982,rhim2012self,Rammal1985,Barrier2020,delplace2010semi,PhysRevB.89.075401,KrishnaKumar181,KrishnaKumar5135}. 
	Inprints of BZfs in terms of periodic oscillations of the conductance in $1/B$ (for fixed $p$), the Brown-Zak oscillations, have been observed up to high temperatures \cite{KrishnaKumar181,KrishnaKumar5135}.
	
		\begin{figure}[t!]
	    \includegraphics[width=3.4in]{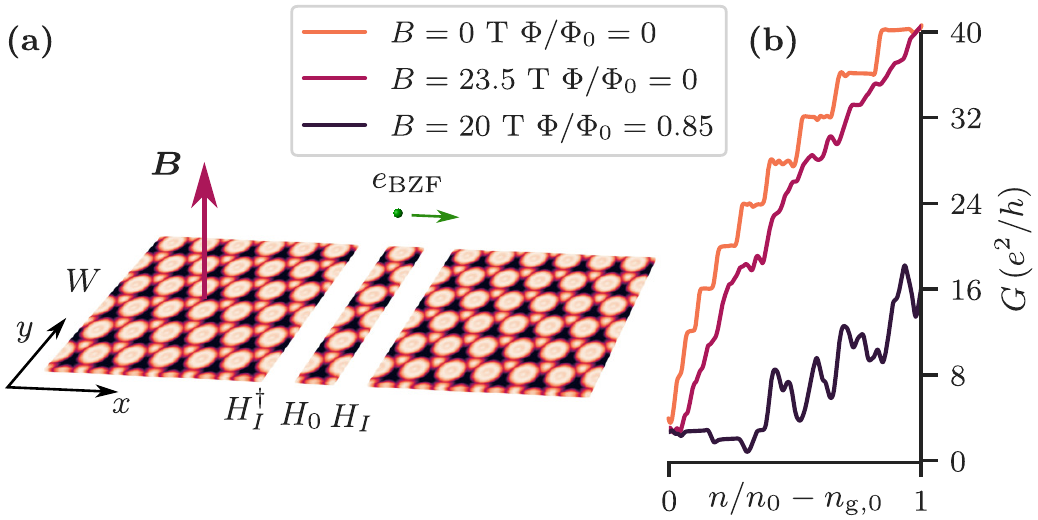}
	    \caption{(a) Build-up of the graphene nanoribbon, schematically. A ``slice'' of one moir\'e cell wide described by $H_0$ is joined by the interaction $H_I$ with adjacent slices to the right and left along the propagation direction of the Brown-Zak Fermion (BZf) $x$. The width $W$ of the nanoribbon is in $y$-direction. The magnetic field is oriented perpendicular to the plane of the ribbon. 
	    (b) Conductance traces for various magnetic fields as a function of normalized charge carrier density $n/n_0$, shifted by the position of the first gap $n_{\text{g},0}$}
	    
	    \label{fSchematic}
	\end{figure}
	
	The large supercell of graphene aligned on hexagonal boron nitride with a linear dimension of $a_S \approx 13.6$ nm includes tenthousands of atoms, rendering numerical simulations of the conductance challenging. 
	Up to now, calculations therefore have focused on the density of states \cite{PhysRevB.90.165404, PhysRevB.89.205414, Jung2015}, a continuum $k\cdot p$ Hamiltonian \cite{PhysRevB.89.075401,Chen2017}, scaled graphene\cite{PhysRevLett.114.036601,Chen2020}, or on effective 1D models\cite{2106.11328}.
	Here we present, to our knowledge, the first realistic large scale simulation of the conductance for a 250~nm wide graphene on hBN nanoribbon. 
	Our algorithm allows treatment of this multi-scale problem extending from the single atomic scale via the scale of the moir\'e supercell $a_S$ (= 13~nm) to the width of the nanoribbon $W=250$~nm for a wide range of magnetic fields and quasiparticle energies.
	Even in the absence of any many-body effects we find a remarkably intricate conductance map which reveals many features observed in recent experiments \cite{Ponomarenko2013,Dean2013,Hunt1427,Yu2014,Wang1231,KrishnaKumar181,KrishnaKumar5135,Barrier2020}.
	This conductance map can be satisfactorily analyzed by a modified Wannier diagram that features properties of both Dirac-like and Schr\"odinger-like quasiparticles.
	
	We simulate a ribbon of aligned ($\Theta=0$) graphene on hexagonal boron nitride in a tight-binding (TB) approximation
		\begin{equation}
			H = \sum_{i,j} t_{ij} c_i c_j^\dagger + \sum_i V_i c_i c_i^\dagger + g_s \mu_B \hat S \cdot \hat B,
		\end{equation}
	with hoppings $t_{ij}$, on site-potential $V_i$ and a Zeeman term with $g_s=2$ and Bohr magneton $\mu_B$.
	The valley degree of freedom of graphene is implicitly contained in the full TB treatment of the lattice and the spin degree of freedom enters through the Zeeman term. 
	We first parametrize the configuration space of the atomistic moir\'e lattice from a set of ab-initio DFT calculation of primitive cells\cite{PhysRevB.84.035440,Quan2021}. 
	We then use a mechanical elasticity model to account for the structural geometry relaxation and induced strain on the graphene layer by the hBN substrate \cite{Nam2017,Carr2018} to build an ab-initio derived graphene/hBN structure (see supplement).
	Unlike most previous theoretical works, we do not use periodic boundary conditions but treat a 250~nm wide ribbon  with zig-zag edges locally covered by a Berry-Mondragon potential to suppress edge-states\cite{doi:10.1098/rspa.1987.0080}. 
	Since we use a general Bloch ansatz in $x$-direction, our setup allows us to compute the conductance for arbitrary magnetic fields. 
	
	\begin{figure*}[t!]
		\includegraphics[width=2.3in]{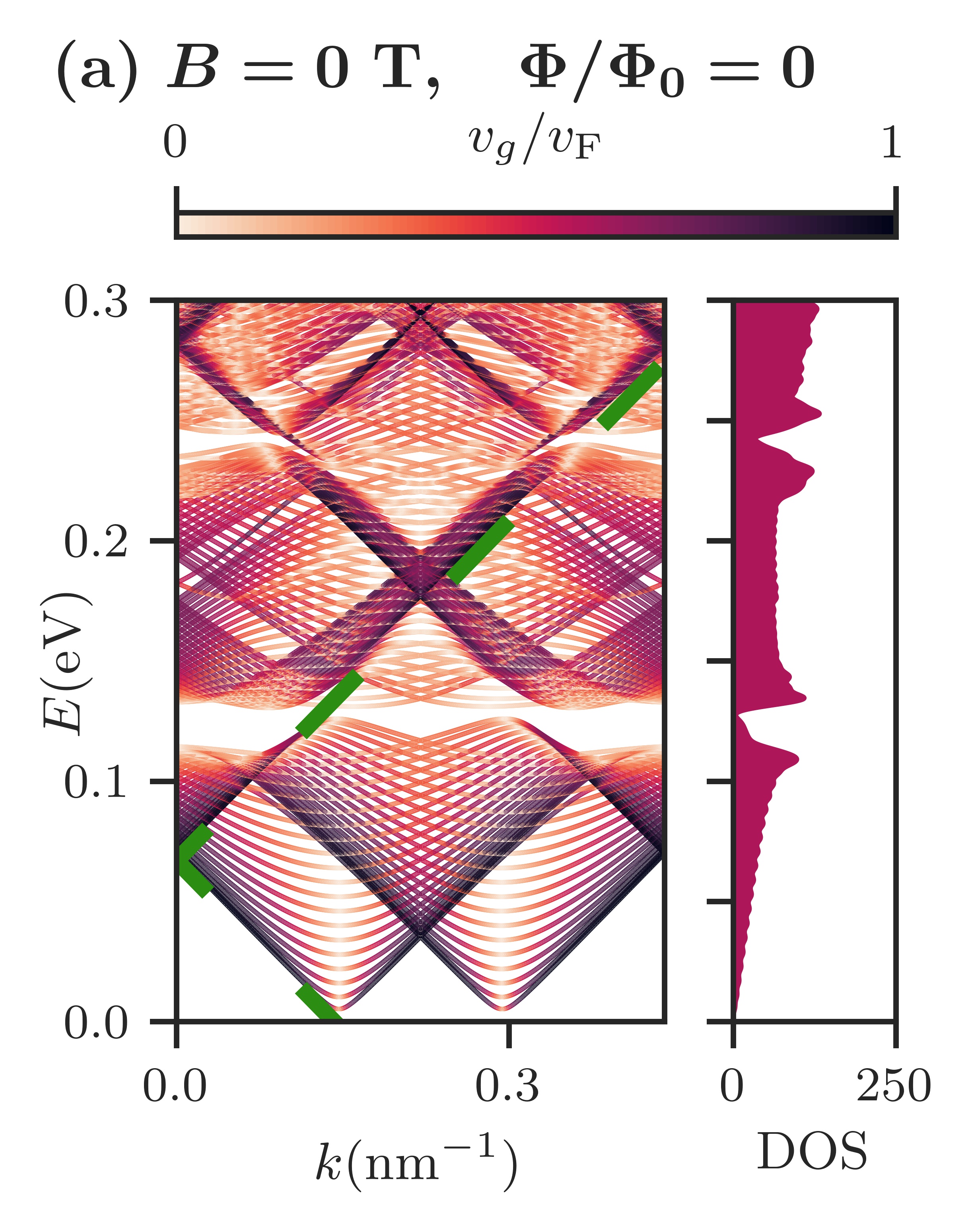}
		\includegraphics[width=2.3in]{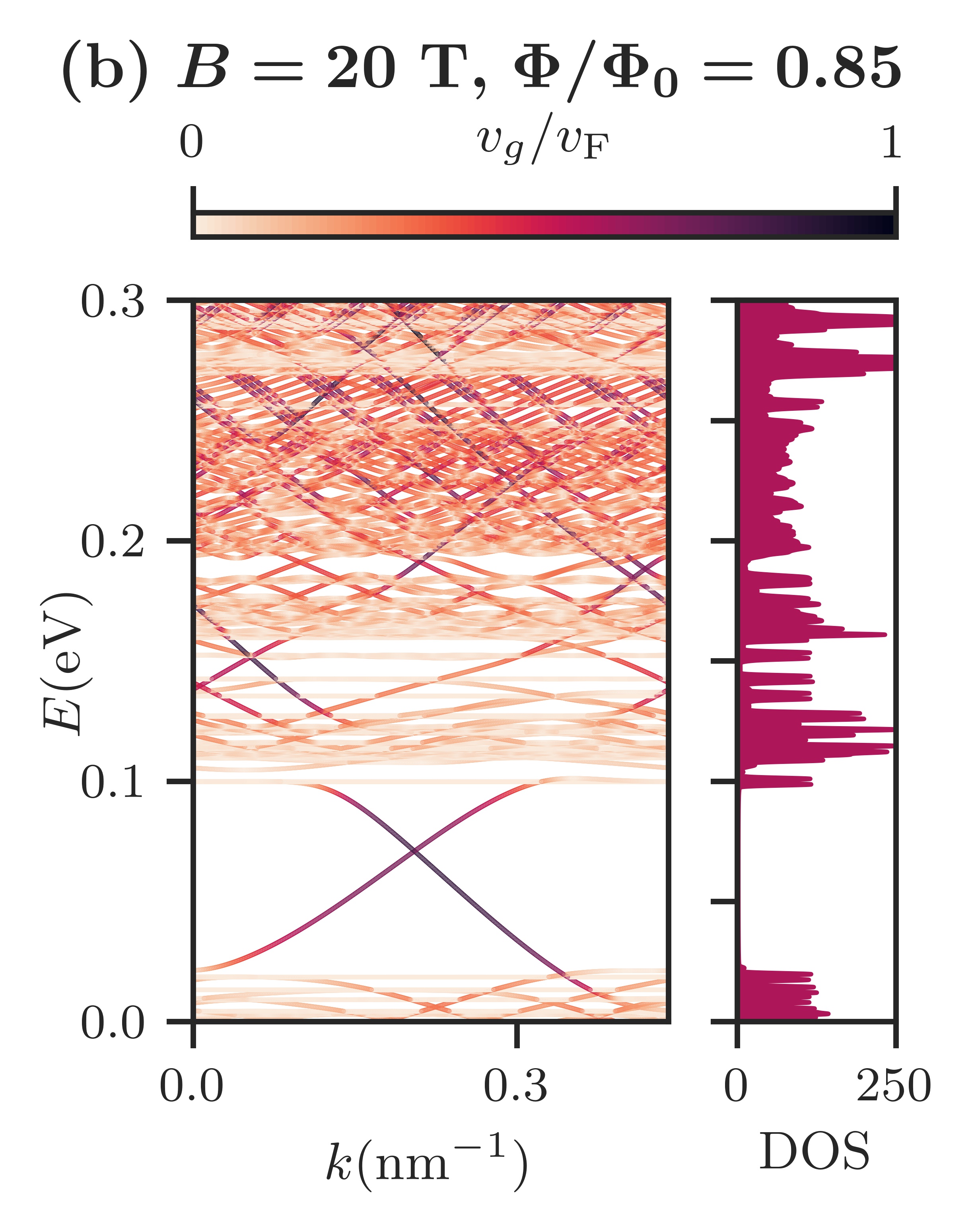}
		\includegraphics[width=2.3in]{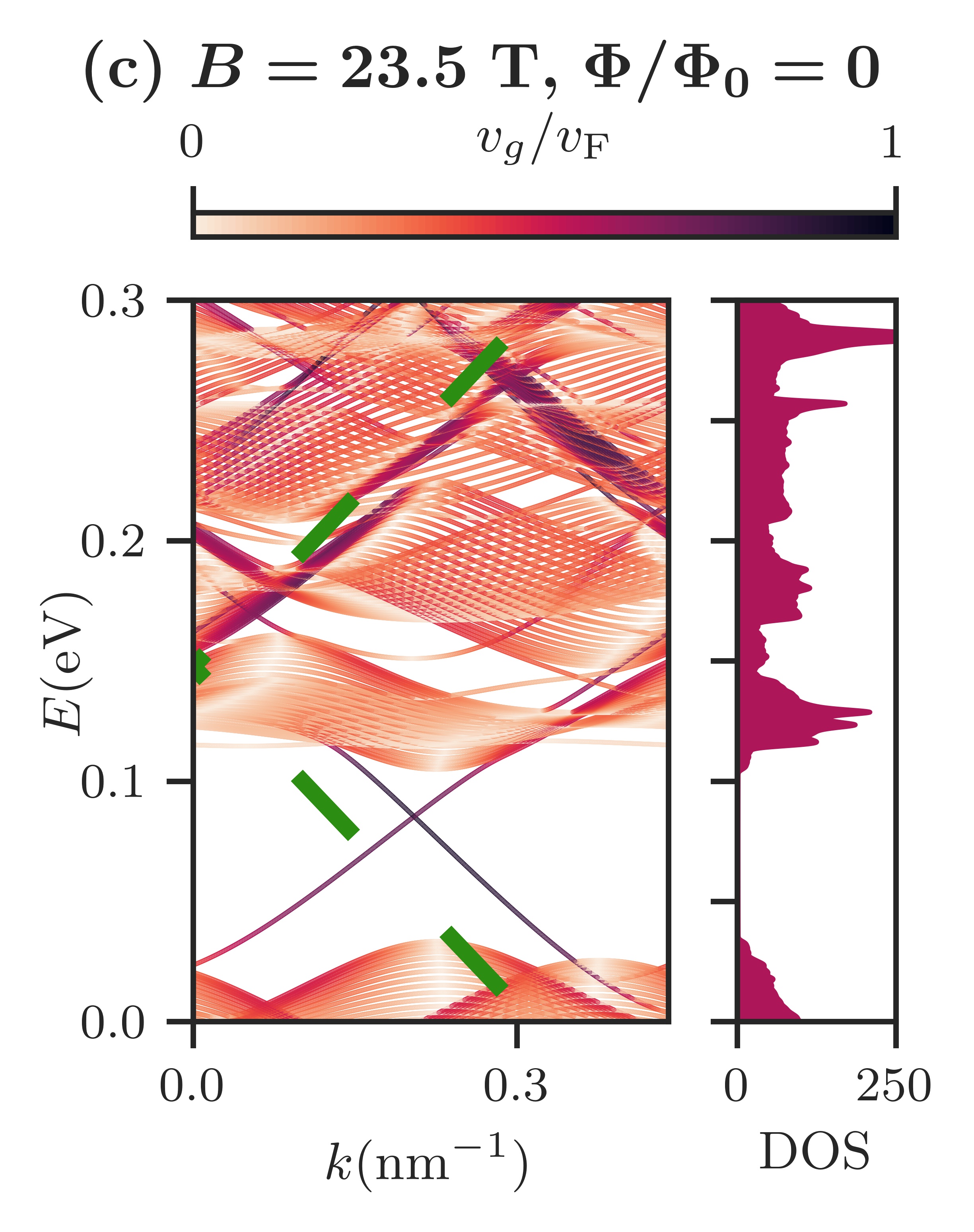}
		\caption{
			Band structure and density of states (DOS) in units $n/n_0/eV$ for the magnetic fields \textbf{(a)} $B=0$~T ($\Phi/\Phi_0=0$), \textbf{(b)} $B=20$~T ($\Phi/\Phi_0 = 0.851$), \textbf{(c)} $B=23.5$~T ($\Phi/\Phi_0=1$). The bandstructure is shaded according to the group velocity relative to $v_{\mathrm{F}}$, the group velocity of the massless Dirac fermions in $(a)$. The slope $v_g = v_{\mathrm{F}}$ is also indicated in (c) by the green dashed line, where it is shifted according to the minimal coupling $\hbar k - e A$ with $A$ the vector potential.
		}
		\label{fBandstruct}
	\end{figure*}
	
	We compute the bandstructure $E(k)$ of the ribbon for various magnetic fields from
		\begin{equation}
			\left(	H_0 + \mathrm{e}^{\mathrm i k \Delta x} H_I + \mathrm{e}^{-\mathrm i k \Delta x} H_I^\dagger \right) \psi_n = E_n(k) \, \psi_n,
			\label{eBandstruct}
		\end{equation} 	
	where $H_0$ is the Hamiltonian of a ``slice'' of the ribbon with  a width in $y$-direction of $20$  moiré unit cells containing approximately two million sites and $H_I$ the interaction Hamiltonian between adjacent slices (Fig.~\ref{fSchematic}a).
	The wavevector $k \equiv k_x$ points in propagation direction.
		To efficiently solve the large but sparse eigenvalue problem  of Eq.~(\ref{eBandstruct}) for eigenvalues $E_n(k)$ and eigenvectors $\psi_n$ close to the Fermi edge, we use the Lanczos method\cite{Lanczos1950AnIM,GOLUB200035} as implemented by the ARPACK library\cite{Arpack}.
		
	As our simulation contains the full information on the band structure, the energy axis can be readily transformed into the charge carrier density simply by counting bands. We set $n=0$ at the charge neutrality point $E=0$ of graphene and express $n$ in units of $n_0 = 1/S$, the density of one electron per moir\'e cell. Likewise, the magnetic field strength $B$ is conveniently expressed in units of magnetic flux quanta through one moir\'e supercell $\Phi/\Phi_0$.

	For each Bloch eigenstate $\psi_n$ the associated group velocity in propagation direction follows from \cite{PhysRevB.78.035407}
		\begin{equation}
		 	v_g^{(n)} = \frac 1\hbar\frac{\partial E_n (k)}{\partial k} = \frac{\mathrm i\Delta x}{\hbar} \psi_n^\dagger \left(H_I  \mathrm{e}^{\mathrm i k \Delta x} \!-\! \mathrm{e}^{-\mathrm i k \Delta x} H_I^\dagger \right) \psi_n.
		 	\label{evg}
		\end{equation}
To efficiently calculate the conductance as function of the energy $E$, we weight each Bloch state moving in $+x$-direction of the Brillouin zone, Eq.~(\ref{eBandstruct}), by the appropriate group velocity, Eq.~(\ref{evg}).
	This weighted density provides an estimate for the conductance $G(E)$ by approximating the number of modes $M(E)$ at each energy,
	\begin{equation}
	    G(E) = \frac{e^2}{h} M \approx \frac{e^2}{h} \frac{\mathrm d}{ \mathrm dE} \sum_{n: E_n < E} \hbar v_g^{(n)} \, \delta k
		\label{eTransportfromBand}
		\end{equation} 
	with a $k$-point spacing $\delta k$.
		Use of Eq.~(\ref{eTransportfromBand}) greatly reduces the computational cost compared to calculation of $G(E)$ using the Landau-B\"utticker formalism \cite{Landauer1957,ButtikerPhysRevB.38.9375} while yielding nearly identical results (see supplement). 
	Our approach is found to be rather efficient: the characteristic $4 \, e^2/h$ conductance steps of a graphene ribbon at zero field are accurately reproduced [Fig~\ref{fSchematic} (b)]. 
	This {\em size quantization} effect is a consequence of the finite ribbon width $W$ with a subband spacing at $B=0$ of the order of 10~meV. 
	The large-scale band structure at zero field (Fig.~\ref{fBandstruct}a) features the Dirac cones with linear dispersion emanating from the K and K' points within which the subbands due to transverse size quantization appear. 
	The prominent gap at about $E_{SD} \approx \pi/a_S v_g \approx 120$ meV is caused by the interaction of the primary cones with the backfolded satellite Dirac (SD) cones generated by the moir\'e potential the strength of which also controls its width $\Delta_m \approx 20$~meV.
	For a generic strong magnetic field ($B=20$~T, or $\Phi/\Phi_0 = 0.851 $) chosen not to coincide with a low-order rational $B_{p/q}$, the conductance is dramatically suppressed  compared to the field-free case (Fig.~\ref{fBandstruct}b), which can be directly traced to the multitude of flat Landau level bands with small group velocities, also visible as pronounced sharp peaks in the density of states (Fig.~\ref{fBandstruct}b). 
	However, a further increase of the magnetic field to $B=23.5$ T corresponding to $B_{p/q} = B_1$ (or $\Phi = \Phi_0$) leads to an almost complete recovery of the zero-field conductance (Fig.~\ref{fSchematic}c). 
	The associated band structure (Fig.~\ref{fBandstruct}c) locally somewhat resembles the zero-field band structure. 
	Most notably, a clustering of Landau subbands with linear slopes close to the slope of the original Dirac cone $v_g \approx v_{\mathrm{F}}$ is observed. 
	It signals the appearance of Brown-Zak fermions (BZfs) when magneto-translational invariance is restored for rational fractions of the flux quantum. 
	The width in energy of the segments over which the linear dispersion of BZfs is visible remains, however, limited by the strength of the moir\'e potential of the order of $20$ meV.
	This perturbative effect of the moir\'e potential is fundamentally different from BZf spectra in pristine lattices where the zero-field band structure (Fig.~\ref{fBandstruct}a) is fully recovered.
    At large $B$, the large gap between the lowest Landau level $n=0$ and $n = \pm 1$, $\Delta_{0,\pm 1}^{(L)} \approx 100$ meV filled only by two quantum Hall edge states, one for each propagation direction develops Figs.~\ref{fBandstruct}(b,c). 
    Smaller gaps remain clearly visible for higher-lying Landau levels (Fig.~\ref{fBandstruct}c). 
    
    \begin{figure}
        \centering
        \includegraphics[width=3.4in]{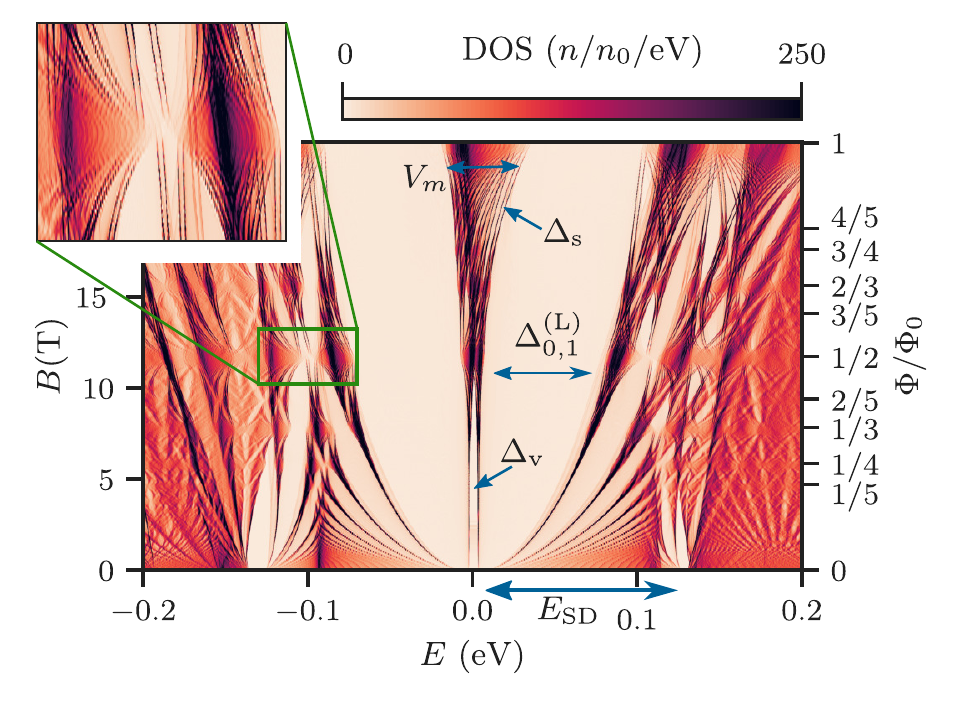}
        \caption{Density of states (DOS) map as a function of $E$ and $B$ featuring multiple energy scales
        $E_{SD} \approx \pi/a_S v_g $: spacing between main Dirac cone and moir\'e induced replica,
        $\Delta_{n-1,n}^{(L)}$: Landau gaps, $\Delta_v$: broken valley degeneracy, $\Delta_s$: spin splitting, $V_m$: Broadening near rational $\Phi/\Phi_0$ by moir\'e potential,
        Inset: zoom into the intersection between Landau level $t=-1$ and the $\Phi/\Phi_0 = 1/2$ line, displaying a Hofstadter butterfly.
        }
        \label{fDOS}
    \end{figure}
    
    The density of states (DOS) map (Fig.~\ref{fDOS}) in the $E-B$ plane not only features the prominent Landau gaps easily recognized by the $\sqrt{B}$ dependence of Landau levels of Dirac particles, but also the presence of multiple energy scales in the presence of both the moir\'e potential and the magnetic field: the gap $E_{\mathrm{SD}} \approx 120$ meV due to the moir\'e-induced secondary Dirac cones, the ``broadening'' of the Landau levels by $\lesssim 20$ meV
    at rational $\Phi/\Phi_0$ (see, e.g., $\Phi/\Phi_0 = 1/2$) caused by the formation of the Hofstadter butterfly delimited by the amplitude $V_m$ of the moir\'e potential, the lifting of the valley degeneracy $\Delta_V \approx 10$ meV by the moir\'e potential and of the spin degeneracy $\Delta_S$ ($\lesssim 3$ meV at $\Phi=\Phi_0$) by the Zeeman term.
    
    \begin{figure*}
        \centering
        \includegraphics[width=7in]{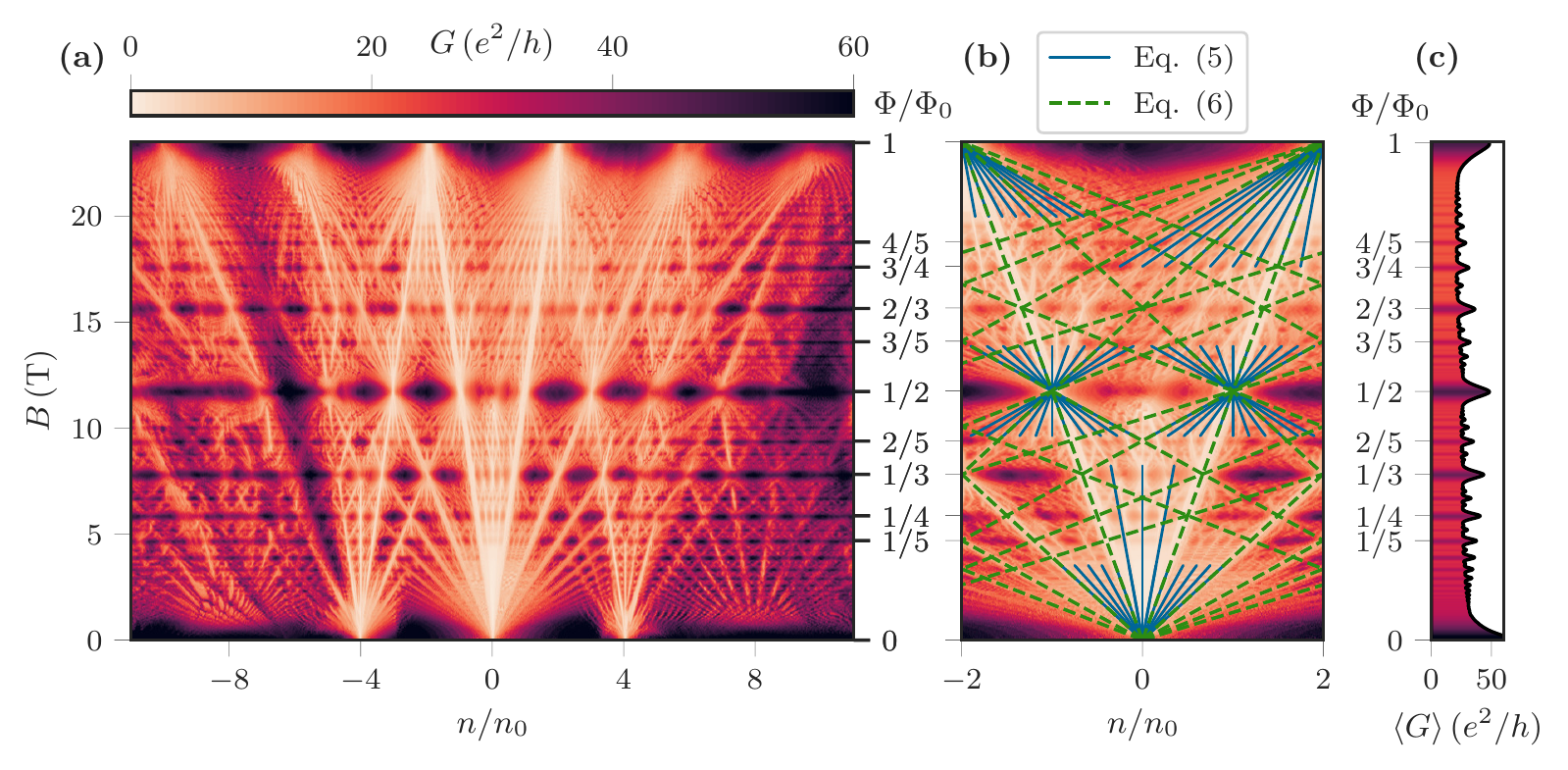}
        \caption{(a) Conductance map $G(n/n_0,B)$ of a 250~nm wide graphene nanoribbon on hBN.
        Additional vertical axis $\Phi/\Phi_0$ denotes the normalized flux per moir\'e unit cell.
        (b) Zoom in of (a), solid blue lines parametrized by the Schrödiner-like Diophantine equation Eq.~(\ref{e:diophantineS}); dashed green lines according to the Dirac Diophantine equation Eq.~(\ref{eWannierGraphene}) for Landau gaps.
        (c) Conductance from (a) averaged over the carrier axis.}
        \label{fConductance}
    \end{figure*}
    
    The simultaneous presence of multiple energy (and length) scales gives rise to a conductance map of surprising complexity displayed as a function of normalized flux $\Phi/\Phi_0$ and normalized carrier density $n/n_0$ (Fig.~\ref{fConductance}), featuring several lines of low and high conductance. 
    Lines of high conductance are strictly horizontal at fixed $B$ and are prominent in the projection on the $B$ (or $\Phi/\Phi_0$) axis.
    $G(B)$ is clearly dominated by a sequence of peaks at rational fractions $p/q$ of $\Phi/\Phi_0$ and is periodic in $1/B$, the so-called Brown-Zak oscillations \cite{PhysRev.133.A1038,PhysRev.134.A1602,PhysRev.134.A1607}. 
    Each conductance peak is associated with the ballistic transport of a BZf. 
    The width of each peak is controlled by the effective field $B_{\mathrm{eff}} = B -B_{p/q}$. 
    Ballistic transport is suppressed when the effective magnetic length $\lambda_{B_{\mathrm{eff}}} = \sqrt{\hbar /(e B_{\mathrm{eff}})}$ 
becomes smaller than the ribbon width $\lambda_{B_{\mathrm{eff}}} \ll W$. 
Accordingly, the peaks of the Brown-Zak oscillations become sharper with increasing ribbon width consistent with recent experimental data for a $17 \mu$m wide sample \cite{Barrier2020}. 

The conductance maps also features a multitude of intersecting straight lines of conductance minima, or ``gaps'', with different slopes, widths, and intersections with the $n/n_0$ axis. 
A subset of these gaps extend over the entire range of $0\le \Phi/\Phi_0 \le 1$, while other, less prominent gaps only open for small field regions (see the zoom-in Fig.~\ref{fConductance} b).
The different types of gaps are well described by using both a modified Wannier diagram for the Landau gaps, taking the linear dispersion of graphene into account, and the usual, ``Schr\"odinger-like'' Wannier diagram for lifted degeneracies on a smaller energy scale.

For a system with Schr\"odinger-like quadratic dispersion originally considered by Hofstadter \cite{PhysRevB.14.2239} and Wannier \cite{PhysRevB.14.2239} the Landau level energies evolve as function of $B$ according to $E_t^{(S)}(B) = \hbar \omega_B(t+1/2)$ with $t \in \mathbb{N}$ and $\omega_B=eB/m$ the cyclotron frequency.
Consequently, the gaps between Landau levels $t-1$ and $t$ are centered at $E_{g,t}^{(S)} = \hbar \omega_B t$. 
The normalized charge carrier density up to the Landau gap $t$ follows in form of a Diophantine equation as 
    \begin{equation}
    \frac{n}{n_0} = \int_0^{E_{g,t}^{(S)}}\rho^{(S)}(E) \, \mathrm{d}E + gs = g \left( t\frac{\Phi}{\Phi_0}+s \right) \label{e:diophantineS}
    \end{equation}
with $s$ the number of filled bands at $E_0=0$, $g$ the degeneracy and $\rho(E) = g/(2\pi)$ the constant density of states for fermions with quadratic dispersion in two dimensions. By contrast, the linear dispersion of massless Dirac (D) fermions results in the Landau levels at $E_t^{(D)}(B) = \mathrm{sgn}(t)v_{\mathrm{F}}\sqrt{2\hbar \left|teB\right|}$ with $t\in\mathbb{Z}$. Here, $t=0$ corresponds to the zero-energy Landau level, a peculiarity of Dirac fermions \cite{Zhang2005,Novoselov2005}. Taking into account the linear increase of the density of states $\rho^{(D)}\propto \left|E\right|$, near the Dirac point an analogous calculation immediately yields a modified Diophantine equation for Landau level gaps of graphene near the charge neutrality point 
    \begin{equation}
    		\frac{n}{n_0} = g^{(D)} \left( \left( t+\frac 1 2 \right) \cdot \frac{\Phi}{\Phi_0} +  s \right), \qquad  s, t \in \mathbb Z.
    		\label{eWannierGraphene}
    \end{equation}
with $g^{(D)}$	the degeneracy of the levels on the Dirac cones. For full valley and spin degeneracy $g^{(D)} = 4$. For massless Dirac fermions the minimal degeneracy is $g^{(D)} \ge 2$ due to Fermion doubling, realized as valley degeneracy in graphene, as consequence of the Nielsen–Ninomiya theorem\cite{NIELSEN198120}. 
	Thus, the gaps described by Eq.~(\ref{eWannierGraphene}) are, despite the half-integer slopes $(t+1/2)$, {\em always} a proper subset of those resulting from Eq.~(\ref{e:diophantineS}). 
	
	The present results are fully consistent with those for the Hall conductivity. 
	More generaly, the slopes $t$ of the gaps in magnetic field can be expressed by the Hall conductivity  $\sigma_{xy} =e \partial n / \partial B|_{E=E_F} = t \cdot e^2/h$.~\cite{PhysRevLett.49.405,Streda_1982}. For graphene, however, a characteristic 
``half integer'' Quantum Hall conductance \cite{Zhang2005,Novoselov2005} $\sigma_{xy} = g ( t + 1/2) \, e^2/h$, $t \in \mathbb Z$ has been observed. 

If valley and spin degeneracy are strongly broken, the gap structure locally reverts to that predicted by the conventional Diophantine equation (Eq.~\ref{e:diophantineS}) with $g=1$. 
These additional line segments typically terminate near horizontal lines of BZfs at rational $\Phi/\Phi_0$. 
This is because the energy scale $V_m$ of the moiré potential forming the BZfs
(Fig.~\ref{fBandstruct}c) is large compared to the valley degeneracy lifting $\Delta V$
(and spin splittings $\Delta S$) generating these gaps.
Indeed, the moiré potential is responsible for both inducing a
valley degeneracy breaking ($\Delta V$) as well as closing the gaps
near rational $\Phi/\Phi_0$ by strongly perturbing the Landau levels (Fig.~\ref{fDOS}).
The Wannier formula Eq.~(\ref{e:diophantineS}) is thus only applicable for small intervals of the magnetic field while the Dirac-like Landau gaps Eq.~(\ref{eWannierGraphene}) remain visible over the full magnetic field range. Such a complex pattern in the Wannier diagram has been, indeed, observed in recent experiments \cite{Ponomarenko2013,Dean2013,Hunt1427,Yu2014,Wang1231,KrishnaKumar181,KrishnaKumar5135,Barrier2020}. We emphasize that such complexity emerges from a realistic simulation of graphene on hBN within a single-particle description on the DFT level without invoking any many-body effects. The energy scale for the latter, $\Delta_{\text{MB}}$, can be estimated from the flat bands in ``magic angle twisted bilayer graphene''\cite{Cao2018,Yankowitz1059} and from measurements on bilayer graphene quantum dots\cite{2106.08405} to be $\Delta_{\text{MB}} \lesssim 1$ meV and, thus, smaller than the energy scales shaping the structures in the Wannier diagram (Fig.~\ref{fConductance}) in the present case. 

One further important structural insight emerging from the Diophantine equation for Dirac fermions (Eq.~\ref{eWannierGraphene}) is that the Wannier diagram is, unlike Eq.~(\ref{e:diophantineS}), and the one for the pristine honeycomb lattice \cite{Rammal1985}, not periodic in $\Phi_0$. The lines emerging from $n/n_0 = 0,\pm 4$ at $\Phi=0$ reach $n/n_0 = \pm 2,\pm 6$ at $\Phi = \Phi_0$. Conversely, lines that would reach $n/n_0 = 0,\pm 4$ at $\Phi = \Phi_0$ are missing. Several experimental data sets 
	\cite{Ponomarenko2013,Dean2013,Hunt1427,Yu2014,Wang1231,KrishnaKumar181,KrishnaKumar5135,Barrier2020}
	display, indeed, such an asymmetry which has, to the best of our knowledge, not yet been discussed or explained. 
	
	In conclusion, our simulations show in detail the emergence of Brown-Zak fermions in moir\'e superlattices. We have calculated the magnetoconductance for ab-initio derived, relaxed graphene on hexagonal boron nitride, displaying an intricate fractal pattern caused by Dirac fermions and their satellite cone structures. This pattern is described by a Diophantine equation for Landau gaps of massless Dirac particles. Our single-particle simulation includes lifting of valley and spin degeneracy resulting in a fine structure that locally follows the Diophantine equation for Schr\"odinger-like particles. The resulting complex Wannier diagram found in agreement with available experimental data can serve as base line to identify possible additional many-body effects not yet included within the single-particle description. We have also clearly demonstrated that the Wannier diagram for Dirac-like particles is non periodic in $\Phi_0$. Such assymetries have been seen in recent experiments but not yet discussed and analyzed.

	\section{Acknowledgements}
	We thank C.~Stampfer for helpful discussions.
	The computational results presented have been achieved using the Vienna Scientific Cluster (VSC4).

\end{document}